\documentclass[sigconf, screen=true, review=false, anonymous=false, authordraft=False, language=german, language=english]{acmart}
\usepackage{blindtext}
\usepackage{tikz}

\AtBeginDocument{%
  }

\setcopyright{rightsretained}

\setcctype{by}

\acmConference[MuC'25]{Mensch und Computer 2025 – Workshopband, Gesellschaft für Informatik e.V.}{31. August - 03. September 2025}{Chemnitz, Germany}
\acmDOI{10.18420/muc2025-mci-ws13-231}

\begin{document}

\title{Too Immersive for the Field? Addressing Safety Risks in \\ Extended Reality User Studies} 


\settopmatter{authorsperrow=4}

\author{Tanja Koji\'c}
\affiliation{%
 \institution{Quality and Usability Lab, TU Berlin}
\city{Berlin}
\country{Germany}}
\email{tanja.kojich@tu-berlin.de}

\author{Sara Srebot}
\affiliation{%
 \institution{University of Zagreb Faculty of Electrical Engineering and Computing}
\city{Zagreb}
\country{Croatia}}
\email{sara.srebot@fer.hr}

\author{Maurizio Vergari}
\affiliation{%
  \institution{Quality and Usability Lab, TU Berlin}
  \city{Berlin}
  \country{Germany}
}
\email{maurizio.vergari@tu-berlin.de}

\author{Mirta Moslavac}
\affiliation{%
 \institution{University of Zagreb Faculty of Electrical Engineering and Computing}
\city{Zagreb}
\country{Croatia}}
\email{mirta.moslavac@fer.hr}

 \author{Maximilian Warsinke}
\affiliation{%
  \institution{Quality and Usability Lab, TU Berlin}
  \city{Berlin}
  \country{Germany}}
\email{warsinke@tu-berlin.de}

\author{Sebastian M\"oller}
\affiliation{%
  \institution{Quality and Usability Lab, TU Berlin \& DFKI}
  \city{Berlin}
  \country{Germany}
}
\email{sebastian.moeller@tu-berlin.de}

\author{Lea Skorin-Kapov}
\affiliation{%
 \institution{University of Zagreb Faculty of Electrical Engineering and Computing}
\city{Zagreb}
\country{Croatia}}
\email{lea.skorin-kapov@fer.hr}

\author{\hbox{Jan-Niklas Voigt-Antons}}
\affiliation{%
  \institution{Immersive Reality Lab, \hbox{Hochschule Hamm-Lippstadt}}
  \city{Lippstadt}
  \country{Germany}
}
\email{Jan-Niklas.Voigt-Antons@hshl.de}

\renewcommand{\shortauthors}{Koji\'c et al.}

\begin{abstract}
Extended Reality (XR) technologies are increasingly tested outside the lab, in homes, schools, and public spaces. While this shift enables more realistic user insights, it also introduces safety challenges that are often overlooked. Physical risks, psychological distress, and accessibility issues can be increased in field studies and unsupervised testing, such as at home or crowdsourced trials. Without clear instructions, safety decisions are left to individual researchers, raising questions of responsibility and consistency. This position paper outlines key safety risks in XR user testing beyond the lab and calls for practical strategies that are needed to help researchers run XR studies in a safe and inclusive way across different environments.
\end{abstract}




\maketitle

\newcommand\copyrighttext{%
    \footnotesize \textcopyright 2026 IEEE. Personal use
    of this material is permitted. Permission from IEEE
    must be obtained for all other uses, in any current or
    future media, including reprinting/republishing this
    material for advertising or promotional purposes,
    creating new collective works, for resale or
    redistribution to servers or lists, or reuse of any
    copyrighted component of this work in other works.
    https://doi.org/10.18420/muc2025-mci-ws13-231}

\newcommand\copyrightnotice{%
\begin{tikzpicture}[remember picture,overlay,shift=
    {(current page.south)}]
    \node[anchor=south,yshift=10pt] at (0,0)
    {\fbox{\parbox{\dimexpr\textwidth-\fboxsep-
    \fboxrule\relax}{\copyrighttext}}};
\end{tikzpicture}%
}
\copyrightnotice

\section{Introduction and Motivation}
Extended Reality (XR) technologies are increasingly used outside controlled lab environments, with user testing taking place in homes, schools, and public spaces. These real world settings introduce unpredictable safety challenges that are rarely addressed in current research protocols. 
As XR becomes more accessible and research moves into the field, safety appears as a critical aspect of user testing. Meaning safety is essential not only for participant well-being but also for ensuring data quality, inclusivity, and ethical standards \cite{slater2021beyond, madary2016real}. 

While safety is acknowledged as a relevant concern in XR research, it often lacks systematic attention, particularly in studies conducted outside laboratory settings. In these cases, researchers must manage safety across diverse and dynamic environments with limited control over physical conditions. This becomes even more complex in unsupervised studies, such as at home trials or crowdsourced XR testing, where participants interact with immersive content without researcher oversight \cite{ratcliffe2021extended}.
Without established guidelines or institutional support tailored to XR specific risks, responsibility for participant safety often falls entirely on individual researchers, raising questions about accountability, feasibility, and consistency across studies.
This raises a central question: who is responsible for identifying, communicating, and mitigating safety risks in XR user studies conducted outside the lab?

\section{Safety Challenges in XR User Studies}

To better understand the implications of field based user testing in XR, it is important to identify the types of safety risks participants may encounter. Unlike controlled laboratory settings, real world environments introduce a wide range of variables that can compromise both user well-being and study outcomes. These risks are not only physical but also psychological, sensory, and accessibilit related with the potential to affect participants differently depending on context, individual traits, and the nature of the XR content. 

Physical safety is one of the most visible concerns in XR usage. Injury data (e.g., NEISS) show a 352\% increase in VR related incidents between 2013 and 2021 \cite{cucher2023virtual}, with fractures, lacerations, and strains among the most common. Reported cases include collisions with furniture, tripping over cables, and more serious injuries during physically demanding games \cite{baur2021cervical}. Such risks are especially relevant in field settings, where physical spaces vary and safety boundaries are often improvised.

Beyond physical harm, psychological distress is another significant risk, especially when immersive content brings out intense emotional responses. Applications that simulate horror, heights, or confined spaces can trigger anxiety or panic \cite{lin2017fear}. Some users have reported trauma like symptoms following unexpected or overwhelming virtual scenarios, particularly in unsupervised contexts where support or debriefing is not available.
Sensory overload can also pose serious challenges for some users. Fast moving visual effects, bright lighting, and intense audio can provoke migraines, trigger epileptic seizures, or overwhelm users with sensory processing sensitivities \cite{cobb1999virtual}. This is especially important to consider in inclusive research designs, as neurodivergent users may be disproportionately affected by these stimuli.

Lastly, many XR systems present accessibility barriers that limit participation or compromise safety for users with physical or sensory impairments. Common issues include the lack of seated play options, limited locomotion settings (e.g., teleportation), and non customizable input controls \cite{mott2019accessible}. These design limitations can exclude participants with disabilities or those recovering from injury and can increase physical strain during testing, especially in longer or more active sessions.

\section{Current Practices and Limitations}
Safety in XR user studies is rarely addressed through standardized procedures. While some platforms provide general warnings—such as the risk of motion sickness or seizures, these are typically generic and not tailored to individual applications \cite{stanney2021extended}. In most cases, comfort ratings rely on developer self assessment, and users or researchers have limited ability to filter content based on safety relevant criteria \cite{vlahovic2023initiative}.

In academic research, safety is usually covered under general ethics protocols, but these often lack guidance specific to the unique risks posed by immersive technologies \cite{slater2021beyond}. Informed consent forms tend to focus on data protection and privacy, while physical, psychological, and sensory risks, such as falling, fatigue, or emotional distress—are not consistently addressed in sufficient detail.

Although a few researchers have begun to share informal safety guidelines or document adverse effects in publications, these efforts remain fragmented. For example, recent work has proposed enhanced rating systems that include comfort, accessibility, and safety dimensions to support more transparent reporting and informed user decisions \cite{vlahovic2023initiative}. However, there is still no widely adopted framework for assessing, documenting, or communicating XRspecific risks across studies. As user testing continues to move beyond the lab, more coordinated and transparent approaches are needed to ensure safe and ethical research practices in varied environments.

\subsection{Why Current Research Practices Fall Short}

While lab based XR studies typically follow institutional safety protocols, field studies present a different set of challenges. In these settings, researchers often rely on improvised setups and personal equipment, which vary widely in quality, calibration, and safety provisions. This variability increases the likelihood of overlooked risks, especially when testing takes place in uncontrolled or unfamiliar environments.

Commercial platforms attempt to provide guidance through comfort labels like “Moderate” or “Intense,” yet these classifications are vague and lack standardized definitions. Because ratings are usually based on developer self assessment, users and researchers alike have limited ability to evaluate or compare safety related features across applications \cite{vlahovic2023initiative}.
The issue extends to research protocols as well. While informed consent forms typically address data privacy and storage, they rarely reflect the specific physical, psychological, or sensory risks associated with XR use \cite{kourtesis2024comprehensive, slater2020ethics}. 
Without standards or centralized resources to guide safety planning in XR user studies, researchers are left to make independent judgments \cite{sudhakaran2024stepping}.

\section{Conclusion}

Field based XR studies provide valuable insights into real-world use, but safety must be recognized as a central concern in their design and execution. 
As immersive research continues to move beyond the lab, there is a growing need for clearer expectations, shared standards, and coordinated responsibilities. Researchers cannot address these challenges alone. Developers, platforms, and institutions must collaborate to support safe, ethical, and inclusive XR research practices across different environments.

\bibliographystyle{ACM-Reference-Format}
\bibliography{main}


\end{document}